\documentclass[12pt]{article}
\begin{document}
\title{
\vspace*{-40pt} CPT symmetry and properties of the exact\\ and
approximate effective Hamiltonians\\ for the neutral kaon
complex\footnote{Talk given at the conference on {\em "Symmetry in
Nonlinear Mathematical Physics"}, June 23 --- 29, 2003, Institute
of Mathematics, National Academy of Sciences of Ukraine, Kiev,
Ukraine.} }
\author{ \hfill \\ Krzysztof URBANOWSKI\footnote{
e--mail: K.Urbanowski@proton.if.uz.zgora.pl;
K.Urbanowski@if.uz.zgora.pl}
\\
University of Zielona G\'{o}ra, Institute of Physics, \\
ul. Podg\'{o}rna 50, 65-246 Zielona G\'{o}ra, Poland.} \maketitle

\begin{abstract}
We start from a discussion of  the general form and general CP--
and CPT-- transformation properties of the Lee--Oehme--Yang (LOY)
effective Hamiltonian for  the neutral kaon complex. Next we show
that there exists an approximation which is more accurate than the
LOY, and which leads to an effective Hamiltonian whose diagonal
matrix elements posses CPT transformation properties, which differ
from those of the LOY effective Hamiltonian. These properties of
the mentioned effective Hamiltonians are compared with the
properties of the exact effective Hamiltonian for the neutral kaon
complex. We show that the diagonal matrix elements of the exact
effective Hamiltonian governing the time evolution in the subspace
of states of an unstable particle and its antiparticle need not be
equal at for $t
> t_{0}$ ($t_{0}$ is the instant of creation of the pair) when the
total system under consideration is CPT invariant but CP
noninvariant. The unusual consequence of this result is that,
contrary to the properties of stable particles, the masses of the
unstable particle "1" and its antiparticle "2" need not be equal
for $t \gg t_{0}$ in the case of preserved CPT and violated CP
symmetries.
\end{abstract}

\section{Introduction.}

The problem of testing CPT--invariance experimentally has
attracted the attention of physicist, practically since the
discovery of antiparticles. CPT symmetry is a fundamental theorem
of axiomatic quantum field theory which follows from locality,
Lorentz invariance, and unitarity \cite{cpt}. Many tests of
CPT--invariance consist in searching for decay process of neutral
kaons. All known CP--  and hypothetically  possible CPT--violation
effects in neutral kaon complex are  described by solving the
Schr\"{o}dinger--like evolution equation \cite{Lee1} ---
\cite{improved} (we use $\hbar = c = 1$ units)
\begin{equation}
i \frac{\partial}{\partial t} |\psi ; t >_{\parallel} =
H_{\parallel} |\psi ; t >_{\parallel} \label{l1}
\end{equation}
for $|\psi ; t >_{\parallel}$ belonging to the subspace ${\cal
H}_{\parallel} \subset {\cal H}$ (where ${\cal H}$ is the state
space of the physical system under investigation), e.g., spanned
by orthonormal neutral  kaons states $|K_{0}>, \;
|{\overline{K}}_{0}>$, and so on, (then states corresponding with
the decay products belong to ${\cal H} \ominus {\cal
H}_{\parallel} \stackrel{\rm def}{=} {\cal H}_{\perp}$), and
nonhermitean effective Hamiltonian $H_{\parallel}$ obtained
usually by means of  Lee-Oehme--Yang (LOY) approach (within the
Weisskopf--Wigner approximation (WW)) \cite{Lee1} ---
\cite{Comins}, \cite{improved}:
\begin{equation}
H_{\parallel} \equiv M - \frac{i}{2} \Gamma, \label{new1}
\end{equation}
where
\begin{equation}
M = M^{+}, \; \; \Gamma = {\Gamma}^{+}, \label{new1a}
\end{equation}
are $(2 \times 2)$ matrices.

Solutions of Eq. (\ref{l1}) can be written in matrix  form  and
such  a  matrix  defines  the evolution    operator (which    is
usually     nonunitary) $U_{\parallel}(t)$ acting in ${\cal
H}_{\parallel}$:
\begin{equation}
|\psi ; t >_{\parallel} = U_{\parallel}(t) |\psi ;t_{0} = 0
>_{\parallel} \stackrel{\rm def}{=} U_{\parallel}(t) |\psi
>_{\parallel}, \label{l1a}
\end{equation}
where,
\begin{equation}
|\psi >_{\parallel} \equiv q_{1}|{\bf 1}> + q_{2}|{\bf 2}>,
\label{l1b}
\end{equation}
and $|{\bf 1}>$ stands for  the  vectors of the   $|K_{0}>,  \;
|B_{0}>$ type and $|{\bf 2}>$ denotes antiparticles  of  the
particle "1": $|{\overline{K}}_{0}>, \; {\overline{B}}_{0}>$,
$<{\bf j}|{\bf k}> = {\delta}_{jk}$, $j,k =1,2$.

In many papers it is assumed that the real parts, $\Re (.)$, of
the diagonal matrix elements of $H_{\parallel}$:
\begin{equation}
\Re \, (h_{jj} ) \equiv M_{jj}, \; \;(j =1,2), \label{m-jj}
\end{equation}
where
\begin{equation}
h_{jk}  =  <{\bf j}|H_{\parallel}|{\bf k}>, \; (j,k=1,2),
\label{h-jk}
\end{equation}
correspond to the masses  of particle "1" and its antiparticle "2"
respectively \cite{Lee1} --- \cite{improved}, (and such an
interpretation of $\Re \, (h_{11})$ and $\Re \, (h_{22})$ will be
used in this paper), whereas the imaginary parts, $\Im (.)$,
\begin{equation}
-2 \Im \, (h_{jj}) \equiv {\Gamma}_{jj}, \; \;(j =1,2),
\label{g-jj}
\end{equation}
are interpreted as the decay widths of these particles \cite{Lee1}
--- \cite{improved}. Such an interpretation seems to be consistent
with the recent and the early experimental data for neutral kaon
and similar complexes \cite{data}.

Relations between matrix elements of  $H_{\parallel}$  implied  by
CPT--transformation properties of the Hamiltonian $H$ of the total
system, containing  neutral  kaon  complex  as  a  subsystem,  are
crucial for designing CPT--invariance and CP--violation tests  and
for proper interpretation of their results. The aim of this paper
is to examine the properties of the approximate and exact
$H_{\parallel}$ generated by the CPT--symmetry of the total system
under consideration and independent of the approximation used.

\section{ {$H_{LOY}$} and CPT--symmetry.}

Now, let us consider briefly some properties of the LOY model. Let
$H$ be total (selfadjoint) Hamiltonian, acting in $\cal H$  ---
then  the total unitary evolution  operator $U(t)$  fulfills  the
Schr\"{o}dinger equation
\begin{equation}
i \frac{\partial}{\partial t} U(t)|\phi > = H U(t)|\phi >,  \; \;
U(0) = I,\label{Schrod}
\end{equation}
where $I$ is the unit operator in $\cal H$, $|\phi > \equiv |\phi
; t_{0} = 0> \in {\cal H}$  is  the  initial  state  of  the
system:
\begin{equation}
|\phi  >  \equiv  |\psi  >_{\parallel}  \label{l2a}
\end{equation}
In  our case  $U(t) |\phi > = |\phi ;t>$. Let $P$ denote the
projection operator onto the subspace ${\cal H}_{\parallel}$:
\begin{equation}
P{\cal H} = {\cal H}_{\parallel}, \; \; \; P = P^{2} = P^{+},
\label{new2}
\end{equation}
then the subspace of decay products ${\cal H}_{\perp}$ equals
\begin{equation}
{\cal H}_{\perp}  = (I - P) {\cal H} \stackrel{\rm def}{=} Q {\cal
H}, \; \; \; Q \equiv I - P. \label{l7c}
\end{equation}
For the  case of neutral  kaons  or  neutral  $B$--mesons,  etc.,
the projector $P$ can be chosen as follows:
\begin{equation}
P \equiv |{\bf 1}><{\bf 1}| + |{\bf 2}><{\bf 2}|. \label{P}
\end{equation}
We assume that time independent basis vectors $|K_{0}>$ and
$|{\overline{K}}_{0}>$ are defined analogously to corresponding
vectors used in LOY theory of time evolution in neutral kaon
complex \cite{Lee1}. In the LOY approach it is assumed that
vectors $|{\bf 1}>$, $|{\bf 2}>$ considered above are eigenstates
of $H^{(0)}$ for 2-fold degenerate eigenvalue $m_{0}$:
\begin{equation}
H^{(0)} |{\bf j} > = m_{0} |{\bf j }>, \; \;  j = 1,2, \label{b1}
\end{equation}
where $H^{(0)}$ is a so called free Hamiltonian, $H^{(0)} \equiv
H_{strong} = H - H_{W}$, and $H_{W}$ denotes weak and other
interactions which are responsible for transitions between
eigenvectors of $H^{(0)}$, i.e., for the decay process.

This means that
\begin{equation}
[P, H^{(0)}] = 0. \label{new3}
\end{equation}

The condition guaranteeing the occurrence of transitions  between
subspaces ${\cal H}_{\parallel}$ and ${\cal H}_{\perp}$, i.e.,  a
decay process of states in ${\cal H}_{\parallel}$, can  be written
as follows
\begin{equation}
[P,H_{W}] \neq 0 , \label{r32}
\end{equation}
that is
\begin{equation}
[P,H] \neq 0 . \label{[P,H]}
\end{equation}
Usually, in LOY and related approaches, it is assumed that
\begin{equation}
{\Theta}H^{(0)}{\Theta}^{-1} = {H^{(0)}}^{+} \equiv H^{(0)} ,
\label{r31}
\end{equation}
where $\Theta$ is the antiunitary operator:
\begin{equation}
\Theta \stackrel{\rm def}{=} {\cal C}{\cal P}{\cal T}.
\label{new4}
\end{equation}
The subspace of neutral kaons ${\cal H}_{\parallel}$ is assumed to
be invariant under $\Theta$:
\begin{equation}
{\Theta} P {\Theta}^{-1} = P^{+} \equiv P. \label{9aa}
\end{equation}

In the kaon rest frame, the time evolution is governed by the
Schr\"{o}dinger equation (\ref{Schrod}), where the initial state
of the system has the form (\ref{l2a}), (\ref{l1b}). Within
assumptions (\ref{b1}) --- (\ref{r32}) the  Weisskopf--Wigner
approach, which is the source of the LOY method, leads to the
following formula for $H_{LOY}$ (e.g., see
\cite{Lee1,Lee2,Cronin,improved}):
\begin{eqnarray}
H_{LOY} = m_{0} P  - \Sigma (m_{0})
& \equiv & PHP - \Sigma (m_{0}), \label{b3} \\
& = & M_{LOY} - \frac{i}{2}{\Gamma}_{LOY} \label{b3a}
\end{eqnarray}
where it has been assumed that $<{\bf 1}|H_{W}|{\bf 2}> = <{\bf
1}|H_{W}|{\bf 2}>^{\ast} = 0$ (see \cite{Lee1} ---
\cite{improved}),
\begin{equation}
\Sigma ( \epsilon ) = PHQ \frac{1}{QHQ - \epsilon - i 0} QHP.
\label{r24}
\end{equation}
The matrix elements $h_{jk}^{LOY}$  of $H_{LOY}$ are
\begin{eqnarray}
h_{jk}^{LOY} & = &  H_{jk} - {\Sigma}_{jk} (m_{0} ) , \; \; \;
(j,k = 1,2) ,  \label{b5} \\
& = & M_{jk}^{LOY} - \frac{i}{2} {\Gamma}_{jk}^{LOY} \label{b5a}
\end{eqnarray}
where, in this case,
\begin{equation}
H_{jk} = <{\bf j} |H| {\bf k} > \equiv <{\bf j} |(H^{(0)} + H_{W}
)| {\bf k} > \equiv m_{0} {\delta}_{jk} + <{\bf j}|H_{W}|{\bf k}>
, \label{b6}
\end{equation}
and ${\Sigma}_{jk} ( \epsilon ) = < {\bf j} \mid \Sigma ( \epsilon
) \mid {\bf k} >$.

Now, if ${\Theta}H_{W}{\Theta}^{-1} = H_{W}^{+} \equiv H_{W}$,
that is if
\begin{equation}
[ \Theta , H] = 0, \label{[CPT,H]=0}
\end{equation}
then using, e.g., the following phase convention \cite{Lee2}
--- \cite{improved}
\begin{equation}
\Theta |{\bf 1}> \stackrel{\rm def}{=} - |{\bf 2}>, \;\;
\Theta|{\bf 2}> \stackrel{\rm def}{=} - |{\bf 1}>, \label{cpt1}
\end{equation}
and taking into account that $< \psi | \varphi > =
<{\Theta}{\varphi}|{\Theta}{\psi}>$, one easily finds from
(\ref{b3}) -- (\ref{b6})  that
\begin{equation}
{h_{11}^{LOY}}^{\Theta} - {h_{22}^{LOY}}^{\Theta} = 0,  \label{b8}
\end{equation}
and thus
\begin{equation}
M_{11}^{LOY} = M_{22}^{LOY}, \label{LOY-m=m}
\end{equation}
(where ${h_{jk}^{LOY}}^{\Theta}$ denotes the matrix elements of
$H_{LOY}^{\Theta}$ --- of the LOY effective Hamiltonian when the
relation (\ref{[CPT,H]=0}) holds), in the CPT--invariant system.
This  is  the standard result of the LOY approach and this is the
picture  which one meets in the literature \cite{Lee1}  ---
\cite{dafne}.

If to assume that CPT--symmetry is not conserved in  the  physical
system under consideration, i.e., that
\begin{equation}
[ \Theta , H] \neq 0, \label{e67}
\end{equation}
then    $h_{11}^{LOY}    \neq    h_{22}^{LOY}$.

It  is  convenient to  express difference between
$H_{LOY}^{\Theta}$   and  the effective Hamiltonian $H_{LOY}$
appearing  within  the  LOY approach in the case  of nonconserved
CPT--symmetry as follows
\begin{eqnarray}
H_{LOY} & \equiv & H_{LOY}^{\Theta} + \delta H_{LOY}  \label{e68} \\
& = & \left(
\begin{array}{cc}
( M_{0} + \frac{1}{2} \delta M) - \frac{i}{2} ( {\Gamma}_{0} +
\frac{1}{2} \delta \Gamma ),
& M_{12} - \frac{i}{2} {\Gamma}_{12} \\
M_{12}^{\ast} - \frac{i}{2} {\Gamma}_{12}^{\ast} & (M_{0} -
\frac{1}{2} \delta M) - \frac{i}{2} ({\Gamma}_{0} - \frac{1}{2}
\delta \Gamma )
\end{array}
\right) .  \nonumber
\end{eqnarray}
In other words
\begin{equation}
h_{jk}^{LOY} = {h_{jk}^{LOY }}^{  \Theta} + \Delta h_{jk}^{LOY},
\label{LOY=h+delta}
\end{equation}
where
\begin{equation}
\Delta h_{jk}^{LOY} = (-1)^{j+1}\frac{1}{2}(\delta M - \frac{i}{2}
\delta \Gamma ) {\delta}_{jk}, \label{LOY-delta}
\end{equation}
and $j,k = 1,2$. Within this approach the $\delta M$ and $\delta
\Gamma$ terms violate CPT--symmetry.

\section{Beyond the LOY approximation}

The more accurate approximate formulae for $H_{\parallel}(t)$ have
been derived in \cite{9,10}  using  the  Krolikowski--Rzewuski
equation for the projection of a state vector \cite{7}, which
results from the Schr\"{o}dinger  equation (\ref{Schrod}) for  the
total system under consideration, and, in the  case  of initial
conditions of the type (\ref{l2a}), takes the following form
\begin{equation}
( i \frac{\partial}{ {\partial} t} - PHP ) U_{\parallel}(t)
 =  - i \int_{0}^{\infty} K(t - \tau ) U_{\parallel}
( \tau ) d \tau,   \label{KR1}
\end{equation}
where $ U_{\parallel} (0)  =  P$,
\begin{equation}
K(t)  =  {\mit \Theta} (t) PHQ \exp (-itQHQ)QHP, \label{K}
\end{equation}
and ${\mit \Theta} (t)  =  { \{ } 1 \;{\rm for} \; t \geq 0, \; \;
0 \; {\rm for} \; t < 0 { \} }$.

The integro--differential equation (\ref{KR1}) can be replaced by
the following differential one (see \cite{bull} --- \cite{7})
\begin{equation}
( i \frac{\partial}{ {\partial} t} - PHP - V_{||}(t) )
U_{\parallel}(t) = 0, \label{KR2}
\end{equation}
where
\begin{equation}
PHP + V_{||} (t) \stackrel{\rm def}{=} H_{||}(t). \label{H||=def}
\end{equation}
Taking into account (\ref{KR1}) and (\ref{KR2}) or (\ref{l1}) one
finds from (\ref{l1a}) and (\ref{KR1})
\begin{equation}
V_{\parallel} (t) U_{\parallel} (t) = - i \int_{0}^{\infty} K(t -
\tau ) U_{\parallel} ( \tau ) d \tau \stackrel{\rm def}{=} - iK
\ast U_{\parallel} (t) . \label{V||=def}
\end{equation}
(Here the asterix, $\ast$, denotes the convolution: $f  \ast g(t)
= \int_{0}^{\infty}\, f(t - \tau ) g( \tau  ) \, d \tau$
).\linebreak Next, using this relation and a retarded Green's
operator  $G(t)$ for the equation (\ref{KR1})
\begin{equation}
G(t) = - i {\mit \Theta} (t) \exp (-itPHP)P, \label{G}
\end{equation}
one obtains \cite{9,10}
\begin{equation}
U_{\parallel}(t) = \Big[ {\it 1} + \sum_{n = 1}^{\infty} (-i)^{n}L
\ast \ldots \ast L \Big] \ast U_{\parallel}^{(0)} (t) ,
\label{U||-szer}
\end{equation}
and thus from (\ref{V||=def})
\begin{equation}
V_{\parallel}(t) \; U_{\parallel}(t) = - i K \ast \Big[ {\it 1} +
\sum_{n = 1}^{\infty} (-i)^{n}L \ast \ldots \ast L \Big] \ast
U_{\parallel}^{(0)} (t) , \label{V-szer})
\end{equation}
where $L$ is convoluted $n$ times, ${\it 1} \equiv {\it 1}(t)
\equiv \delta (t)$,
\begin{equation}
L(t) = G \ast K(t), \label{L=GK}
\end{equation}
\begin{equation}
U_{\parallel}^{(0)} = \exp (-itPHP) \; P \label{U0}
\end{equation}
is a "free" solution of Eq. (\ref{KR1}). Of course, the  series
(\ref{V-szer}) is convergent if \linebreak $\parallel L(t)
\parallel < 1$. If for every $t \geq 0$
\begin{equation}
\parallel L(t) \parallel \ll 1, \label{L<1}
\end{equation}
then, to the lowest order of  $L(t)$,  one  finds  from
(\ref{V-szer}) \cite{9,10}
\begin{equation}
V_{\parallel}(t) \cong V_{\parallel}^{(1)} (t) \stackrel{\rm
def}{=} -i \int_{0}^{\infty} K(t - \tau ) \exp {[} i ( t - \tau )
PHP {]} d \tau . \label{V||=approx}
\end{equation}
Thus \cite{8,pra,9,10}
\begin{equation}
H_{\parallel}(0) \equiv PHP, \; \; V_{\parallel}(0) = 0, \; \;
V_{\parallel} (t \rightarrow 0) \simeq -itPHQHP. \label{H||(0)}
\end{equation}

In the case of (\ref{P}) of the projector $P$ and such $H$ that
\begin{equation}
PHP \equiv m_{0}\, P, \label{P-H12=0}
\end{equation}
that is for
\begin{equation}
H_{12} = H_{21} = 0,  \label{H12=0}
\end{equation}
the operator $P e^{itPHP}$ takes the following form,
\begin{equation}
P e^{\textstyle{ i t PHP}} = P e^{\textstyle{itm_{0}}},
\label{exp-PHP-H}
\end{equation}
and thus the  approximate formula (\ref{V||=approx}) for
$V_{\parallel}(t)$ leads to
\begin{equation}
V_{\parallel}^{(1)} (t) = - PHQ \frac{e^{\textstyle{-it(QHQ -
m_{0})}} - 1}{QHQ - m_{0}} QHP, \label{V||(t)-H0}
\end{equation}
which leads to $V_{||} \stackrel{\rm def}{=} \lim_{t \rightarrow
\infty} V_{||}^{(1)} (t)$,
\begin{equation}
V_{||} = - \Sigma (m_{0}). \label{V-H-0}
\end{equation}
This means that in the case (\ref{P-H12=0})
\begin{equation}
H_{||} = m_{0} \, P - \,\Sigma (m_{0}), \label{H||-H12=0}
\end{equation}
and  $H_{||} = H_{LOY}$.

On the other hand, in the case
\begin{equation}
H_{12} = H_{21}^{\ast} \neq 0, \label{H12n0}
\end{equation}
the form of $Pe^{itPHP}$ is more complicated. For example in the
case of conserved CPT,  formula (\ref{V||=approx})  leads to the
following form for $V_{||} \stackrel{\rm def}{=} \lim_{t
\rightarrow \infty} V_{||}^{(1)} (t)$ \cite{improved,Piskorski}
\begin{eqnarray}
V_{||}^{\mit\Theta} & = & - \frac{1}{2} \Sigma (H_{0} +
|H_{12}|)\, \Big[ \Big( 1 - \frac{H_{0}}{|H{_{12}|}} \Big)P +
\frac{1}{|H_{12}|} PHP \Big]
\nonumber \\
& &  - \frac{1}{2} \Sigma (H_{0} - |H_{12}|)\, \Big[ \Big( 1 +
\frac{H_{0}}{|H{_{12}|}} \Big)P - \frac{1}{|H_{12}|} PHP \Big],
\label{V-H12n0}
\end{eqnarray}
where
\begin{equation}
H_{0} \stackrel{\rm def}{=} \frac{1}{2} ( H_{11} + H_{22} ),
\label{H0}
\end{equation}
and $V_{\parallel}^{\mit \Theta}$ denotes $V_{\parallel}$ when
(\ref{[CPT,H]=0}) occurs.

In the general case (\ref{H12n0}), when there are not any
assumptions about symmetries of the type CP--,  T--, or
CPT--symmetry for the total Hamiltonian  H of the system
considered, the form of  $V_{||} = V_{\parallel}(t \rightarrow
\infty ) \cong V_{\parallel}^{(1)} ( \infty )$ is a yet more
complicated. In such a case one finds the following expressions
for the matrix elements  $v_{jk}(t \rightarrow \infty )
\stackrel{\rm def}{=} v_{jk}$  of $V_{\parallel}$ \cite{9,10},
\begin{eqnarray}
v_{j1} = & - & \frac{1}{2} \Big( 1 + \frac{H_{z}}{\kappa} \Big)
{\Sigma}_{j1} (H_{0} + \kappa ) - \frac{1}{2} \Big( 1 -
\frac{H_{z}}{\kappa} \Big)
{\Sigma}_{j1} (H_{0} - \kappa )\nonumber   \\
& - & \frac{H_{21}}{2 \kappa} {\Sigma}_{j2} (H_{0} + \kappa ) +
\frac{H_{21}}{2 \kappa} {\Sigma}_{j2} (H_{0} - \kappa ) ,
\nonumber \\
&  & \label{v-jk}\\
v_{j2} = & - & \frac{1}{2} \Big( 1 - \frac{H_{z}}{\kappa} \Big)
{\Sigma}_{j2} (H_{0} + \kappa ) - \frac{1}{2} \Big( 1 +
\frac{H_{z}}{\kappa} \Big)
{\Sigma}_{j2} (H_{0} - \kappa ) \nonumber  \\
& - & \frac{H_{12}}{2 \kappa} {\Sigma}_{j1} (H_{0} + \kappa ) +
\frac{H_{12}}{2 \kappa} {\Sigma}_{j1} (H_{0} - \kappa ) ,\nonumber
\end{eqnarray}
where $j,k = 1,2$,
\begin{equation}
H_{z} = \frac{1}{2} ( H_{11} - H_{22} ) , \label{H-z}
\end{equation}
and
\begin{equation}
\kappa = ( |H_{12} |^{2} + H_{z}^{2} )^{1/2} . \label{kappa}
\end{equation}
Hence, by (\ref{H||=def})
\begin{equation}
h_{jk} = H_{jk} + v_{jk}. \label{h-jk-bis}
\end{equation}
It should be emphasized that  all components  of the expressions
(\ref{v-jk}) have the same  order  with respect  to $\Sigma (
\varepsilon )$.

In the case of preserved  CPT--symmetry  (\ref{[CPT,H]=0}),  one
finds $H_{11} = H_{22}$ which implies that $\kappa \equiv |H_{12}
|$, $H_{z} \equiv 0$ and  $H_{0} \equiv H_{11} \equiv H_{22}$, and
\cite{9,10}
\begin{equation}
{\Sigma}_{11} ( \varepsilon = {\varepsilon}^{\ast} ) \equiv
{\Sigma}_{22} ( \varepsilon = {\varepsilon}^{\ast} ) \stackrel{\rm
def}{=} {\Sigma}_{0} ( \varepsilon = {\varepsilon}^{\ast} ) .
\label{sigma-11}
\end{equation}
Therefore  matrix  elements  $v_{jk}^{\mit \Theta}$  of operator
$V_{\parallel}^{\mit \Theta}$  take the following form
\begin{eqnarray}
v_{j1}^{\mit \Theta} = & - & \frac{1}{2} {\Big\{ } {\Sigma}_{j1}
(H_{0} + | H_{12} |)
+ {\Sigma}_{j1} (H_{0} - | H_{12} |)  \nonumber \\
& + & \frac{H_{21}}{|H_{12}|} {\Sigma}_{j2} (H_{0} + | H_{12} |) -
\frac{H_{21}}{|H_{12}|} {\Sigma}_{j2} (H_{0} - | H_{12} |) {\Big\}
} , \nonumber \\
& & \label{v-jk-cpt} \\
v_{j2}^{\mit \Theta} = & - & \frac{1}{2} {\Big\{ } {\Sigma}_{j2}
(H_{0} + | H_{12} |)
+ {\Sigma}_{j2} (H_{0} - | H_{12} |)  \nonumber \\
& + & \frac{H_{12}}{|H_{12}|} {\Sigma}_{j1} (H_{0} + | H_{12} |) -
\frac{H_{12}}{|H_{12}|} {\Sigma}_{j1} (H_{0} - | H_{12} |) {\Big\}
}, \nonumber
\end{eqnarray}

Assuming
\begin{equation}
|H_{12}| \ll |H_{0}|, \label{H<H0}
\end{equation}
we find
\begin{equation}
v_{j1}^{\mit  \Theta} \simeq - {\Sigma}_{j1} (H_{0} ) - H_{21}
\frac{ \partial {\Sigma}_{j2} (x) }{\partial x}
\begin{array}[t]{l} \vline \, \\ \vline \,
{\scriptstyle x = H_{0} } \end{array} , \label{v-j1-cpt<}
\end{equation}
\begin{equation}
v_{j2}^{\mit  \Theta} \simeq - {\Sigma}_{j2} (H_{0} ) - H_{12}
\frac{ \partial {\Sigma}_{j1} (x) }{\partial x}
\begin{array}[t]{l} \vline \, \\ \vline \,
{\scriptstyle x = H_{0} } \end{array} , \label{v-j2-cpt<}
\end{equation}
where $j = 1,2$.  One  should  stress  that  due  to  a presence
of resonance terms, derivatives $\frac{\partial}{\partial x}
{\Sigma}_{jk} (x)$ need not  be  small  and  neither need the
products $H_{jk} \frac{\partial}{\partial x} {\Sigma}_{jk}  (x)$
in  (\ref{v-j1-cpt<}), (\ref{v-j2-cpt<}) .

Finally, assuming that (\ref{H<H0}) holds and using relations
(\ref{v-j1-cpt<}), (\ref{v-j2-cpt<}), (\ref{h-jk}) and the
expression (\ref{b5}), we obtain for the CPT--invariant system
\cite{is,hep-ph-0202253}
\begin{equation}
h_{j1}^{\mit  \Theta} \simeq  h_{j1}^{\rm LOY} - H_{21} \frac{
\partial {\Sigma}_{j2} (x) }{\partial x}
\begin{array}[t]{l} \vline \, \\ \vline \,
{\scriptstyle x = H_{0} } \end{array}  \stackrel{\rm def}{=}
h_{j1}^{\rm LOY} + \delta h_{j1}, \label{h-j1-cpt<}
\end{equation}
\begin{equation}
h_{j2}^{\mit  \Theta} \simeq  h_{j2}^{\rm LOY} - H_{12} \frac{
\partial {\Sigma}_{j1} (x) }{\partial x}
\begin{array}[t]{l}  \vline \, \\ \vline \,
{\scriptstyle x = H_{0} } \end{array}  \stackrel{\rm def}{=}
h_{j2}^{\rm LOY} + \delta h_{j2}, \label{h-j2-cpt<}
\end{equation}
where $j = 1,2$. From these formulae we  conclude  that, e.g., the
difference  between  diagonal   matrix   elements   of
$H_{\parallel}^{\mit \Theta}$, which plays  an  important  role in
designing CPT--invariance tests  for  the  neutral  kaons  system,
equals

\begin{equation}
\Delta h \stackrel{\rm def}{=} h_{11} - h_{22} \simeq H_{12}
\frac{ \partial {\Sigma}_{21} (x) }{\partial x}
\begin{array}[t]{l} \vline \, \\ \vline \,
{\scriptstyle x = H_{0} } \end{array} - H_{21} \frac{ \partial
{\Sigma}_{12} (x) }{\partial x}
\begin{array}[t]{l} \vline \, \\ \vline \,
{\scriptstyle x = H_{0} }\end{array} \neq 0, \label{delta-h}
\end{equation}
which differs from the LOY results (\ref{b8}), (\ref{LOY-m=m}).

\section{CPT and the exact effective Hamiltonian}

The aim of this Section is to show, that contrary to the LOY
conclusion (\ref{b8}), diagonal matrix elements of the exact
effective Hamiltonian $H_{||}$ can not be equal when the total
system under consideration is CPT invariant but CP noninvariant.
This will be done by means of the method used in \cite{plb2002}.

Universal properties of the (unstable) particle--antiparticle
subsystem of the system described by the  Hamiltonian $H$, for
which  the relation (\ref{[CPT,H]=0}) holds, can be extracted from
the matrix elements of the exact $U_{||}(t)$ appearing in
(\ref{l1a}). Such $U_{||}(t)$ has the following form
\begin{equation}
U_{||}(t) = P U(t)P, \label{U||}
\end{equation}
where $P$ is defined by the relation (\ref{P}), and $U(t)$ is the
total unitary evolution operator $U(t)$, which solves the
Schr\"{o}\-din\-ger equation (\ref{Schrod}). Operator $U_{||}(t)$
acts in the subspace of unstable states ${\cal H}_{||} \equiv P
{\cal H}$. Of course, $U_{||}(t)$ has nontrivial form only if
(\ref{[P,H]}) holds, and only then transitions of states from
${\cal H}_{||}$ into ${\cal H}_{\perp}$ and vice versa, i.e.,
decay and regeneration processes, are allowed.

Using the matrix representation one finds
\begin{equation}
U_{||}(t) \equiv \left(
\begin{array}{cc}
{\rm \bf A}(t) & {\rm \bf 0} \\
{\rm \bf 0} & {\rm \bf 0}
\end{array} \right)
\label{A(t)}
\end{equation}
where ${\rm \bf 0}$ denotes the suitable zero submatrices and a
submatrix ${\rm \bf A}(t)$ is the $2 \times 2$ matrix acting in
${\cal H}_{||}$
\begin{equation}
{\rm \bf A}(t) = \left(
\begin{array}{cc}
A_{11}(t) & A_{12}(t) \\
A_{21}(t) & A_{22}(t)
\end{array} \right) \label{A(t)=}
\end{equation}
and $A_{jk}(t) = <{\bf j}|U_{||}(t)|{\bf k}> \equiv <{\bf
j}|U(t)|{\bf k}>$, $(j,k =1,2)$.

Now assuming (\ref{[CPT,H]=0}) and using the phase convention
(\ref{cpt1}), \cite{Lee1} --- \cite{Comins}, one easily finds that
\cite{chiu}, \cite{nowakowski} --- \cite{leonid2}
\begin{equation}
A_{11}(t) = A_{22}(t). \label{A11=A22}
\end{equation}
Note that assumptions (\ref{[CPT,H]=0}) and (\ref{cpt1}) give no
relations between $A_{12}(t)$ and $A_{21}(t)$.

The important relation between amplitudes $A_{12}(t)$ and
$A_{21}(t)$ follows from the famous Khalfin's Theorem \cite{chiu},
\cite{leonid1} --- \cite{leonid2}. This Theorem states that in the
case of unstable states, if amplitudes $A_{12}(t)$ and $A_{21}(t)$
have the same time dependence
\begin{equation}
r(t) \stackrel{\rm def}{=} \frac{A_{12}(t)}{A_{21}(t)} = {\rm
const} \equiv r, \label{r=const},
\end{equation}
then it must be $|r| = 1$.

For unstable particles the relation (\ref{A11=A22}) means that
decay laws
\begin{equation}
p_{j}(t) \stackrel{\rm def}{=} |A_{jj}(t)|^{2}, \label{p-j}
\end{equation}
(where $j = 1,2$), of the particle $|{\bf 1}>$ and its
antiparticle $|{\bf 2}>$ are equal,
\begin{equation}
p_{1}(t) \equiv p_{2}(t). \label{p1=p2}
\end{equation}
The consequence of this last  property is that the decay rates of
the particle $|{\bf 1}>$ and its antiparticle $|{\bf 2}>$ must be
equal too.

From (\ref{A11=A22}) it does not follow that the masses of the
particle "1" and the antiparticle "2" should be equal.

More conclusions about the properties of the matrix elements of
$H_{||}$ one can infer analyzing the following identity
\cite{horwitz}, \cite{bull} --- \cite{pra}
\begin{equation}
H_{||} \equiv H_{||}(t) = i \frac{\partial U_{||}(t)}{\partial t}
[U_{||}(t)]^{-1}, \label{H||2a}
\end{equation}
where $[U_{||}(t)]^{-1}$ is defined as follows
\begin{equation}
U_{||}(t) \, [U_{||}(t)]^{-1} = [U_{||}(t)]^{-1} \, U_{||}(t) \, =
\, P. \label{U^-1}
\end{equation}
(Note that the identity (\ref{H||2a}) holds, independent of
whether $[P,H] \neq 0$ or $[P,H]=0$). The expression (\ref{H||2a})
can be rewritten using the matrix ${\bf A}(t)$

\begin{equation}
H_{||}(t) \equiv  i \frac{\partial {\bf A}(t)}{\partial t} [{\bf
A}(t)]^{-1}. \label{H||2b}
\end{equation}
Relations (\ref{H||2a}), (\ref{H||2b}) must be fulfilled by the
exact as well as by every approximate effective Hamiltonian
governing the time evolution in every two dimensional subspace
${\cal H}_{||}$ of states $\cal H$ \cite{horwitz}, \cite{bull} ---
\cite{pra}.

It is easy to find from (\ref{H||2a}) the general formulae for the
diagonal matrix elements, $h_{jj}$, of $H_{||}(t)$, in which we
are interested. We have
\begin{eqnarray}
h_{11}(t) &=& \frac{i}{\det {\bf A}(t)} \Big( \frac{\partial
A_{11}(t)}{\partial t} A_{22}(t) - \frac{\partial
A_{12}(t)}{\partial t} A_{21}(t) \Big), \label{h11=} \\
h_{22}(t) & = & \frac{i}{\det {\bf A}(t)} \Big( - \frac{\partial
A_{21}(t)}{\partial t} A_{12}(t) + \frac{\partial
A_{22}(t)}{\partial t} A_{11}(t) \Big). \label{h22=}
\end{eqnarray}
Now, assuming (\ref{[CPT,H]=0}) and using the consequence
(\ref{A11=A22}) of this assumption, one finds
\begin{equation}
h_{11}(t) - h_{22}(t) =  \frac{i}{\det {\bf A}(t)} \Big(
\frac{\partial A_{21}(t)}{\partial t} A_{12}(t) - \frac{\partial
A_{12}(t)}{\partial t} A_{21}(t) \Big). \label{h11-h22=}
\end{equation}
Next, after some algebra one obtains
\begin{equation}
h_{11}(t) - h_{22}(t) = - i \, \frac{A_{12}(t) \, A_{21}(t) }{\det
{\bf A}(t)} \; \frac{\partial}{\partial t} \ln
\Big(\frac{A_{12}(t)}{A_{21}(t)} \Big). \label{h11-h22=1}
\end{equation}
This result means that in the considered case for $t>0$ the
following Theorem holds:
\begin{equation}
h_{11}(t) - h_{22}(t) = 0 \; \; \Leftrightarrow \; \;
\frac{A_{12}(t)}{A_{21}(t)}\;\; = \; \; {\rm const.}, \; \; (t >
0). \label{h11-h22=0}
\end{equation}
Thus for $t > 0$ the problem under studies is reduced to the
Khalfin's Theorem (see the relation (\ref{r=const})).

From (\ref{h11=}) and (\ref{h22=}) it is easy to see that at $t=0$
\begin{equation}
h_{jj}(0) = <{\bf j}|H|{\bf j}>, \; \; (j=1,2), \label{hjjt=0}
\end{equation}
which means that in a CPT invariant system (\ref{[CPT,H]=0}) in
the case of pairs of unstable particles, for which transformations
of type (\ref{cpt1}) hold
\begin{equation}
M_{11}(0) = M_{22}(0) \equiv <{\bf 1}|H|{\bf 1}>, \label{M11=M22}
\end{equation}
the unstable particles "1" and "2" are created at $t=t_{0} \equiv
0$ as  particles with equal masses.

Now let us go on to analyze the  conclusions following from the
Khalfin's Theorem. CP noninvariance requires that $|r| \neq 1$
\cite{chiu,nowakowski,leonid1,leonid2} (see also \cite{Lee1}
--- \cite{Cronin}, \cite{data}). This means that in such a case it
must be $r = r(t) \neq {\rm const.}$. So, if in the system
considered the property (\ref{[CPT,H]=0}) holds but
\begin{equation}
[{\cal CP}, H] \neq 0, \label{[CP,H]}
\end{equation}
and the unstable states "1" and "2" are connected by a relation of
type (\ref{cpt1}), then at $t > 0$ it must be $(h_{11}(t) -
h_{22}(t)) \neq 0$ in this system. This means that for $t \gg 0$
it can be $\Re \, (h_{11}(t) - h_{22}(t))\, \neq \,0$ and $\Im
\,(h_{11}(t) - h_{22}(t))\, = \, 0$ or $\Re \, (h_{11}(t) -
h_{22}(t))\, = \, 0 $ and $\Im \,(h_{11}(t) - h_{22}(t))\, \neq \,
0$ or $ \Re \, (h_{11}(t) - h_{22}(t)) \, \neq \, 0$ and  $\Im \,
(h_{11}(t) - h_{22}(t)) \, \neq \, 0$ both.

Let us focus our attention on $\Re \,(h_{11}(t) - h_{22}(t))$.
Following the method used in \cite{11} and using assumption
(\ref{9aa}) and  the   identity (\ref{H||2a}), after some algebra,
one finds \cite{hep--ph/9803376}.
\begin{equation}
[\Theta, H_{\parallel}(t)] = {\cal A}(t) + {\cal B}(t),
\label{l10}
\end{equation}\vspace*{-10pt}
where:
\begin{eqnarray}
{\cal A}(t) \; & = & \; P  [{\Theta},H] U(t) P
\bigl( U_{\parallel}(t) {\bigr)}^{-1} , \label{l11} \\
{\cal B}(t) & \equiv &  { \Big{\{} } PH  \: -  \: H_{\parallel}(t)
P { \Big{\}} } [{\Theta} ,U(t)]  P \bigl(
U_{\parallel}(t){\bigr)}^{-1} \label{l13}
\end{eqnarray}

We observe that ${\cal A}(0) \equiv P[\Theta,H]P$ and ${\cal B}(0)
\equiv 0$. From definitions and general properties of operators
$\cal C$,$\cal P$ and $\cal T$ \cite{Comins,Yu-V} it is known that
${\cal T}U(t{\neq}0) = U_{T}^{+}(t{\neq}0){\cal T}$ $\neq$
$U(t{\neq}0){\cal T}$ (Wigner's definition for $\cal T$ is used),
and thereby ${\Theta}U(t \neq 0) = U_{CPT}^{+}(t \neq 0){\Theta}$
i.e. $[\Theta,U(t  \neq  0)] \neq  0$. So, the component ${\cal
B}(t)$ in (\ref{l10}) is nonzero for $t \neq 0$ and  it  is
obvious that there is a chance for $\Theta$--operator to commute
with the effective   Hamiltonian $H_{\parallel}(t \neq 0)$ only if
$[\Theta,H] \neq 0$. On the  other  hand, the property
$[\Theta,H]\neq 0$ does not imply that $[\Theta, H_{\parallel}(0)]
=  0$  or $[\Theta, H_{\parallel}(0)] \neq 0$. These two
possibilities  are admissible, but if $[\Theta, H] =  0$ then
there  is  only  one possibility: $[\Theta, H_{\parallel}(0)] = 0$
\cite{11}.

From (\ref{l10}) we find
\begin{equation}
\Theta H_{\parallel}(t) \Theta^{-1} - H_{\parallel}(t) \equiv
\bigl( {\cal A}(t) + {\cal B}(t) \bigr) \Theta^{-1}. \label{l16}
\end{equation}
Now, keeping in mind that $|{\bf 2}> \equiv  |\overline{K}_{0}>$
is the antiparticle for $|{\bf 1}> \equiv |K_{0}>$ and that,  by
definition, the  (anti--unitary) $\Theta$--operator  transforms
$|{\bf 1}>$   in  $|{\bf 2}>$  according to formulae (\ref{cpt1}),
we obtain from (\ref{l16})
\begin{equation}
h_{11}(t)^{\ast} - h_{22}(t) = <{\bf 2}| \bigl( {\cal A}(t) +
{\cal B}(t) \bigr) {\Theta}^{-1}|{\bf 2}>, \label{l17}
\end{equation}
Adding expression (\ref{l17}) to its  complex  conjugate  one gets
\begin{equation}
{\Re} \; (h_{11}(t) - h_{22}(t)) = {\Re} \; <{\bf 2}| \bigl( {\cal
A}(t) + {\cal B}(t) \bigr) {\Theta}^{-1}|{\bf 2}>. \label{l18}
\end{equation}

Note that if to replace the the requirement (\ref{[P,H]}) for the
projector $P$ (\ref{P}) by the following one:
\begin{equation}
[P,H] = 0, \label{dd1}
\end{equation}
i.e., if to consider only stationary states instead of unstable
states, then one immediately obtains from (\ref{l11}),
(\ref{l13}):
\begin{eqnarray}
{\cal A}(t) \; & = & \; P  [{\Theta},H] P,  \label{dd2} \\
{\cal B}(t) \; & = & \; 0. \label{dd3}
\end{eqnarray}

Let us assume that the property (\ref{[CPT,H]=0}) holds. For the
stationary states (\ref{dd1}), this assumption, relations
(\ref{dd2}), (\ref{dd3}) and (\ref{l18}) yield ${\Re} \;
(h_{11}(t) - h_{22}(t)) = 0$.

Now let us consider the case of unstable states, i.e., states
$|{\bf 1}>, |{\bf 2}>$, which lead to such projection operator $P$
(\ref{P}) that condition (\ref{[P,H]}) holds. If in this case
(\ref{[CPT,H]=0}) also holds then ${\cal A}(t) \equiv 0$ and thus
$[ {\Theta}, H_{\parallel}(0) ]$ $ = 0$, which is in agreement
with (\ref{M11=M22}) and with an  earlier,  similar  result
\cite{11}. In this case we have ${\Theta}U(t)  =  U^{+}(t)\Theta$,
which gives ${\Theta}U_{\parallel}(t) =$
$U_{\parallel}^{+}(t)\Theta$, ${\Theta}U_{\parallel}^{-1}(t)  =
(U_{\parallel}^{+}(t))^{-1}\Theta$, and
\begin{equation}
[\Theta, U(t)]  =  - 2i \bigl( {\Im} \; U(t) \bigr) \Theta
\label{l20}
\end{equation}
This relation leads to the following result in  the  case  of  the
conserved CPT--symmetry
\begin{equation}
{\cal B}(t)  =  -2i P \Big{\{}  H \: - \: H_{\parallel}(t) \: P
\Big{\}} \bigl( {\rm Im} \: U(t) \bigr) P \bigl(
U_{\parallel}^{+}(t) {\bigr)}^{-1}\Theta \label{l21}
\end{equation}

Formula (\ref{l21}) allow us to conclude that $<{\bf 2}|{\cal
B}(0){\Theta}^{-1} |{\bf 2}> = 0$ and \linebreak ${\Re}\,(<{\bf
2}|{\cal B}(t
> 0) {\Theta}^{-1}|{\bf 2}>) \, \neq 0$, if condition (\ref{[CPT,H]=0})
holds. This means that in this case it must be
${\Re}\,(\,h_{11}(t)\,) \neq {\Re}\,(\,h_{22}(t)\,)$ for $t > 0$.
So, there is no possibility for ${\Re}\,(h_{11})$ to equal
${\Re}\,(h_{22})$ for $t> 0$  in  the  considered case of $P$
fulfilling the condition (\ref{[P,H]}) (i.e., for unstable states)
when CPT--symmetry  is conserved: It must be ${\Re}\,(h_{11}) \neq
{\Re}\,(h_{22})$.

Assuming the LOY interpretation of $\Re \,(h_{jj}(t))$, ($j=1,2$),
one can conclude from the Khalfin's Theorem and from the
properties (\ref{h11-h22=0}), (\ref{l18}), (\ref{l21}) that if
$A_{12}(t)$, $A_{21}(t) \neq 0$ for $t > 0$ and if the total
system considered is CPT--invariant, but CP--noninvariant, then
$M_{11}(t) \neq M_{22}(t)$ for $t >0$, that is, that contrary to
the case of stable particles (the bound states), the masses of the
simultaneously created unstable particle "1" and its antiparticle
"2", which are connected by the relation (\ref{cpt1}), need not be
equal  for $t
>t_{0} =0$.  Of course, such a conclusion contradicts
the standard LOY result (\ref{b8}), (\ref{LOY-m=m}). However, one
should remember that the LOY description of neutral $K$ mesons and
similar complexes is only an approximate one, and that the LOY
approximation is not perfect. On the other hand the relation
(\ref{h11-h22=0}) and the Khalfin's Theorem follow from the basic
principles of the quantum theory and are rigorous. Consequently,
their implications should also be considered as rigorous.

\section{Final remarks}

Note that properties of the more accurate approximation described
in Sec. 3 are consistent with the general properties and
conclusions obtained in Sec. 4 for the exact effective Hamiltonian
--- compare (\ref{H||(0)}) and (\ref{hjjt=0}) and relations
(\ref{h11-h22=0}) with (\ref{delta-h}).

From the result (\ref{delta-h}) it follows that in the case of the
approximate $H_{||}$, $\Delta h = 0$ can be achieved only if
$H_{12}=H_{21} = 0$. This means that if the first order $|\Delta
S| = 2$ interactions are forbidden in the $K_{0},
{\overline{K}}_{0}$ complex then predictions following from the
use of the mentioned more accurate approximation  and from the LOY
theory should lead to the the same masses for $K_{0}$ and for
${\overline{K}}_{0}$. This does not contradict the results of Sec.
3 derived for the exact $H_{||}$: the mass difference is very,
very  small and should arise at higher orders of the more accurate
approximation.

On the other hand from (\ref{delta-h}) it follows that $\Delta h
\neq 0$ if and only if $H_{12} \neq 0$. This means that if
measurable deviations from the LOY predictions concerning the
masses  of, e.g. $K_{0}, {\overline{K}}_{0}$ mesons are ever
detected, then the most plausible interpretation of this result
will be the existence of first order $|\Delta S| = 2$ interactions
in the system considered.

\end{document}